\documentclass{JAC2001}

\newcommand{\EN}{\mbox{$\epsilon_n$}}
\newcommand{\GEV}{\mbox{GeV}}
\newcommand{\EE}{\mbox{${\cal{E}}_e$}}
\begin{document}
\title{\vspace*{-0.cm} \hspace*{11cm}  
{\large\rm BUDKER-INP 2003-15} \\[0mm]
\hspace*{9.8cm}  {\large\rm LC-M-2003-062} \\[0mm]
IS A LASER WIRE A NON-INVASIVE METHOD ?~\thanks{Talk at
    26-th Advanced ICFA Beam Dynamic  Workshop on Nanometre-Size
    Colliding Beams (Nanobeam2002), Lausanne, Switzerland, Sept 2-6,
    2002.}}

\author{V.I. Telnov~\thanks{e:mail telnov@inp.nsk.su} \\
{\it Institute of Nuclear Physics, 630090 Novosibirsk, Russia}\\}

\maketitle

\begin{abstract}
  A tightly focused laser beam (laser wire) is used for measurement of
  transverse electron beam sizes in storage rings and linear
  colliders. It is assumed that the laser beam does nothing with the
  electron beam except  Compton scatterings which happen
  with a rather small probability. In reality,
  electrons crossing the laser beam get kicks (with 100\%
  probability) proportional to the square of the laser field and inversely
  proportional to the beam energy.  In  practical cases of beam
  diagnostics  this effect is negligible.
\end{abstract}

\section{INTRODUCTION}
A laser beam (wire) is used  for measurement of electron beam sizes at
storage rings and linear colliders~[1-6].
In this method, a laser beam with a diameter of about one wavelength moves
across the electron beam and Compton scattered photons are
detected. Typically, the probability of Compton scattering
per one electron is about $10^{-7}$. There is a common belief that only
scattered electrons are lost and nothing happens with the other
electrons which cross the laser beam, though they oscillate inside the
laser field, but the resulting kick is zero. 

On the other hand, the diameter of the laser beam is about $\lambda$,
the laser field $E_L$ during the electron crossing is almost unipolar
and one can expect the kick $P_{\perp}c \sim eE_L\lambda$.~\footnote{In
  this paper we omit  all numerical coefficients.}
 Using the  standard formula for the Gaussian
laser beam and performing integration of the Lorentz force along the
unperturbed  electron trajectory one would also get a non-zero result,
though somewhat lower than  given above, but again $P_{\perp} \propto
E_L$. 
     
Below we show that the kick is non-zero and depends on the laser field
quadratically.
\begin{equation}
P_{\perp}c \sim \frac{mc^2 k}{\alpha \gamma} \propto E_L^2,
\end{equation}
where $k$ is Compton scattering probability, $m$ the electron mass,
$c$ the speed of light,  $\gamma = \EE/mc^2$, \EE\ is the electron
energy, $\alpha=e^2/\hbar c = 1/137$ and all numerical coefficients
are omitted. In most practical applications of the laser wire this
effect can be neglected, except for low energy beams and large
conversion coefficients.

Physics of this non-zero result is very interesting and instructive.
This subject was actively discussed during the last decade in physics
community working on laser acceleration of electrons  and
this discussion is still continued. 

\section{Interaction of electrons with a laser beam}

Below we consider electron-laser interactions by two methods that
allow us to understand better the origin of a non-zero energy-momentum
exchange between   electrons and  laser beams.

\subsection{Considerations based on conservation of the energy}
 
The electron energy and momentum can change only if something happens
with the laser beam (beside  Compton scattering where all is clear).
The electron static field  is the same before
and after crossing the laser beam. If the laser field is also unchanged then
the electron kinetic energy should be conserved. Let us take
now into account the electron radiation field. Integrating the
energy density of the electromagnetic field  before and after the
interaction one gets a net change of the electron energy: 
$$\Delta \EE\ \sim \int{({\vec{E}_L +  \vec{E}_r})^2 dV} - \int{E_L^2 dV}$$
\begin{equation}
\;\;\;\; \approx   \int{ (2 { \vec{E}_L \vec{E}_r} + E_r^2) dV},    
\end{equation} 
where $E_L$ is the laser field, $E_r$ the electron radiation field
which interferes with the laser field.   

In the presence of matter near the interaction
region, the electron can radiate without  laser
beam. In this case $E_r$ is  independent of the laser field and
therefore  $\Delta \EE\ \propto E_r E_L \propto E_L $.  Namely
this mechanism explains the energy gain in linear accelerators where
the accelerating gradient is just proportional to the RF  field.
After acceleration of the electron bunch, the field in the linac
becomes weaker. The electron bunch radiates to many cavity modes,  but
the mode which coincides with the accelerator field gives the main
contribution due to the interference of a rather weak radiated field
with a strong RF accelerator field. Depending on the phase it may be
acceleration or deceleration of the beam.  All other radiated modes
just give some small energy loss $\Delta \EE\ \propto E_r^2$.
This picture is well known in the accelerator community.

In the case of the open space  we are interested in, the
electron radiates only due to the acceleration in the laser
field and  therefore $E_r \propto E_L$ and $\Delta \EE\
\propto E_L^2$.

Let us estimate $\Delta \EE$ for the case when the laser beam
diameter is about one wavelength and the electron intersects it
perpendicular. The radiation field in the dipole approximation at
large angles (only such radiation can interfere with the laser beam)
is equal 
\begin{equation}  
E_r \sim \frac{ea}{rc^2}, \;\;\; a \sim \frac {eE_L}{\gamma m}\;\;\; 
 \Rightarrow \;\;\;
E_r \sim \frac{e^2 E_L}{\gamma mc^2 r}\,,
\end{equation}
where $a$ is the electron acceleration and $r$ the distance from the electron.
As the laser is focused on the spot with a diameter $\sim \lambda$ the
characteristic volume is $\Delta V \sim \lambda^3$, $r \sim
\lambda$. As result we get
\begin{equation}  
\Delta \EE\ \sim E_L E_r \Delta V \sim  \frac{e^2 E_L^2
  \lambda^2}{mc^2\gamma} \sim \frac{mc^2 \xi^2}{\gamma},
\label{dE1}
\end{equation}
where 
\begin{equation}
\xi^2= \frac{e^2\langle E_L^2 \rangle}{m^2c^2\omega^2}
\label{xi2}
\end{equation}
is the parameter characterizing the nonlinear effects in  Compton
scattering (or the undulator parameters in wigglers).  The  energy
exchange between the electron and the laser wave $\Delta \EE$
corresponds to absorption of laser photons and emission to the laser
wave  without Compton scattering. Due to the laser divergence,
absorption and emission of photons with different directions may
result in  the kick of the electron in the direction perpendicular to the
laser beam. Therefore (\ref{dE1}) is also the estimate of the
transverse momentum kick $P_{\perp}$.

A free electron can not absorb or emit one  photon but virtual
absorption  and reemission (to some other
direction) is allowed and gives a net kick. The two-photon nature of the
considered effect is seen from the quadratic dependence of the force
on the field strength.   

\subsection{Action of a ponderomotive force}

Let us consider the same effect in a different way, in the language of
ponderomotive forces. In the non-uniform laser field the electron
undergoes fast oscillations and drifts to the region with the lower field.
The latter can be understood in the following way. The energy of the
relativistic electron oscillating  in a weak laser field is~\footnote{For
simplicity we assume that the laser electrical field is perpendicular
to the electron direction, a general case gives similar result.}   
\begin{equation}
\EE\ \sim \sqrt{P^2 c^2 + P_x^2 c^2 + m^2c^4} \sim \EE_0 +
\frac{P_x^2 c^2}{2\EE_0}, 
\end{equation}
where $P_x = (eE_L/\omega) \cos{\omega t}\,\, $, $\omega$ is the laser
frequency.  Substituting $P_x$ and averaging the energy over the fast
oscillations we get
\begin{equation}
\EE\ = \EE_0 + \frac{mc^2 \xi^2}{2\gamma}. 
\end{equation}
The second term can be considered as the potential energy. The
corresponding force is the well known  ponderomotive
force~\cite{kibble,mora} which pushes out the electron from the laser
field
\begin{equation}
F_p = \frac{mc^2}{2\gamma} \nabla{\xi^2}.   
\label{Fp}
\end{equation}
Taking $\nabla{\xi^2} \sim \xi^2/r$ (r is the radius of the laser
beam) we find the kick 
\begin{equation}
\Delta P_{\perp} c \sim F_p r \sim  \frac{mc^2 \xi^2}{\gamma}.
\label{dE2}
\end{equation}
This estimate coincides with (\ref{dE1}) obtained in a completely
different way. It is more transparent and in addition gives a well
defined direction of the force. Eqs.(\ref{Fp}),(\ref{dE2}) are valid
for $r \gg\ \lambda$ (for many oscillations) but can be used 
down to $r\sim \lambda$ as an estimate.

Considering the ponderomotive force we did not mention radiation
fields. They are hidden in the fast oscillations of the electron.  If one
would neglect the electron motion under influence of the laser field,
then the integration of the Lorentz force along the straight electron
trajectory would give a zero result. 

The non-zero result  for the Gaussian laser beams mentioned in the
beginning of the paper is connected with the fact that a Gaussian
description of the field near the laser focus is just a paraxial
approximation corresponding to the case when the laser diameter is
much larger than the wavelength. In the general case of large
diffraction angles, the Gaussian description is not valid because it
does not obey the Maxwell equations. There are papers where high order
corrections to the Gaussian beams are found~\cite{Salamin}.   But even
without formulas it is clear that integration  of the force along any
trajectory (ignoring electron motion in the field) can give only a
result which is proportional to the field strength (because the
Lorentz force is proportional to  the field strength). But the linear
term in the case of free space  is forbidden by simple energy
considerations arguments discussed above.  This statement is known as
the Lawson-Woodward theorem~\cite{lawson}.

\section{Estimation of the electron kick by the laser wire}

Let us express (\ref{dE1},\ref{dE2}) in terms of  the Compton
scattering probability (or the conversion coefficient $k$) which is
\begin{equation}
k= n_{\gamma} \sigma_c l \sim \frac{e^4 E_L^2}{\hbar\omega^2m^2c^3},
\label{k}
\end{equation}
where $n_{\gamma} \sim E_L^2 /(\hbar \omega)$ is the density of laser
photons, $\sigma_c \sim r_e^2 = e^4/m^2c^4$  is the Compton cross
section, $l \sim \lambda$ is the diameter of the laser beam. Using
(\ref{xi2},\ref{k}) we can rewrite (\ref{dE2}) as
\begin{equation}
\Delta P_{\perp} c \sim \frac{kmc^2}{\alpha\gamma}.
\label{dP}
\end{equation} 

This kick should be compared with the rms transverse momenta of electrons in
the beam which are by definition
\begin{equation}
P_{\perp} c =  mc^2 \sqrt{\frac{\EN\gamma}{\beta}},
\label{P}
\end{equation}
where $\EN$ is the normalized emittance and $\beta$ the beta function. The
relative increase of the transverse momentum spread is
\begin{equation}
\frac{\Delta P_{\perp}}{P_{\perp}} \sim \frac{k}{\alpha \gamma}
\sqrt{\frac{\beta}{\EN\gamma}}.
\end{equation}
For $\EN\sim 10^{-6}$ cm (the minimum vertical normalized emittance
considered at damping rings for linear colliders), $\gamma \sim 5
\times 10^{3}$ (damping ring), $\beta = 300 \sqrt{\EE(\GEV)}\,\,\mbox{cm} \sim
500$ cm, $k \sim 10^{-7}$ we get
\begin{equation}
\frac{\Delta P_{\perp}}{P_{\perp}} \sim 10^{-6},   
\end{equation}  
which is negligible.

\section{Conclusion}

  Electrons passing the laser wire do not only undergo
Compton scattering  (with a small probability), but in addition all electrons
receive  transverse kicks with 100\% probability . Fortunately, these kicks
are rather small  and can be ignored in most applications. 

\section*{Acknowledgements}

I would like to thank Karsten Buesser and Frank Zimmermann for reading
the manuscript and useful remarks. This work was supported in part 
by INTAS 00-00679.

\end{document}